\documentclass[floatfix,citeautoscript,nofootinbib,superscriptaddress,twocolumn,pra]{revtex4-2}
\usepackage{amsbsy}
\usepackage{latexsym,epsfig,graphicx}
\usepackage{dcolumn}
\usepackage{graphicx}
\usepackage{color}
\usepackage{bm}
\usepackage{mathrsfs}
\usepackage{amsmath}
\usepackage{amssymb}

\begin{document}
\title{Time-reversed Young's experiment: Deterministic, diffractionless second-order interference effect}
\author{Jianming Wen}
\email{jwen7@binghamton.edu}
\affiliation{Department of Electrical and Computer Engineering, Binghamton University, Binghamton, New York 13902, USA}

\begin{abstract}
The classic Young's double-slit experiment exhibits first-order interference, producing alternating bright and dark fringes shaped by the diffraction effect of the slits. In contrast, here we demonstrate that its time-reversed configuration generates an ideal, deterministic second-order `ghost' interference pattern free from diffraction and first-order effects with the use of only a position-fixed detector. The pattern's size is governed by the dimensions of the `effectively extended light source' formed by point emitter(s). Beyond highlighting the nonreciprocal physics between the two configurations, this system unlocks a range of novel phenomena inaccessible in traditional double-slit experiments. These include fully programmable, digitized interference fringe formations and the ability to align the pattern plane with the source plane on the same side of the setup. Remarkably, our proposed experiment achieves these outcomes without relying on nonclassical correlations or quantum entanglement. By restoring time-reversal symmetry and eliminating diffraction, our theoretical analysis reveals that this approach offers exciting potential for advancing optical imaging and sensing technologies with improved transverse and longitudinal phase-shift sensitivity and resolution limit beyond current limitations.
\end{abstract}

\maketitle

\section{Introduction}
Young’s double-slit experiment stands as a cornerstone in modern physics, bearing profound implications for our comprehension of the nature of light and matter. Originally conducted by Thomas Young \cite{1} in the early nineteenth century, this seminal experiment provided compelling evidence for the wave-like attributes of light through the observation of interference patterns produced by light traversing two closely spaced slits. This groundbreaking revelation challenged the prevalent notion of light solely as a particle and laid the foundation for the wave-particle duality \cite{2,3,4,5,6,7,8,9,10,11,12} concept—a keystone principle of quantum mechanics. Moreover, Young’s experiment elucidated the principles of superposition and coherence, bedrock tenets underpinning various domains of modern physics, including quantum mechanics and optics. Its significance transcends the realm of light, as similar interference phenomena have been observed with matter waves \cite{13,14,15,16,17,18,19,20,21,22,23,24}, reinforcing the unified nature of physics and complementarity interpretation of quantum mechanics. Thus, Young’s interference experiment remains indispensable in molding our understanding of the fundamental principles governing the behavior of light and matter, with far-reaching implications across diverse fields of scientific inquiry.

On the other hand, the diffraction effect assumes a pivotal role in Young’s experiment, broadening our comprehension of wave behavior while simultaneously posing challenges in experimental precision. Positively, diffraction is intrinsic to the creation of the interference pattern observed in the experiment. As light passes through an aperture, it diffracts and spreads out into a succession of wavefronts. According to the Huygens-Fresnel principle, these wavefronts superpose and interfere with each other, engendering regions of constructive and destructive interference, which form the characteristic bright and dark fringes on the screen. Meanwhile, diffraction can present challenges in experimental setup and interpretation. The dispersion of light due to diffraction can obscure the interference pattern, diminishing the sharpness of the fringes and complicating measurements. Additionally, diffraction around the edges of the slits introduces high order maxima and minima, thereby confining the observable scale of the interference pattern. Despite these challenges, understanding and accounting for the diffraction effect are essential for accurately interpreting and applying the results of any traditional Young’s experiment, whether using classical or quantum light sources.

Here, we show that replacing the point light source in the standard double-slit experiment with a position-fixed point (or bucket) detector, and substituting the observation plane with a spatially extended point-light-emitter source, reveals a novel time-reversed configuration that uncovers a variety of intriguing phenomena characterized by counterintuitive effects unattainable in the classic Young's experiment. Notably, this includes the emergence of diffraction-free, deterministic ``ghost" interference fringes formed by the position-fixed detector, where the size of the pattern is dictated by the lateral dimensions of the entire light source. Moreover, this interference formation is fully predictable, programmable, and digitized--capabilities that are impossible in traditional setups. A distinctive feature of this configuration is its departure from conventional physics: the underlying phenomenon is fundamentally tied to the ``two-particle" \textit{second-order correlation} effect, despite the use of only a single position-fixed point detector. This sharply contrasts with the traditional single-particle picture developed from single-photon or electron Young's experiments.

Furthermore, we present a unified theoretical analysis of the spatial and longitudinal sensitivity and resolution limits in this time-reversed Young's experimental configuration. By exploring the phase response to emitter displacement in both lateral (transverse) and axial (longitudinal) directions, we derive the sensitivity functions and resolution limits governed by the fundamental shot-noise constraints. Numerical evaluations based on experimentally relevant parameters are included to provide quantitative benchmarks and to highlight sub-wavelength resolving capability of the scheme.
We anticipate that these findings will not only deepen our understanding of the iconic double-slit experiment but also pave the way for transformative advancements in superresolution imaging and sensing technologies in the post-diffraction era. 

\section{Brief Overview of Classic Young's Experiment}
\begin{figure}[htbp]
\centering\includegraphics[width=0.9\linewidth]{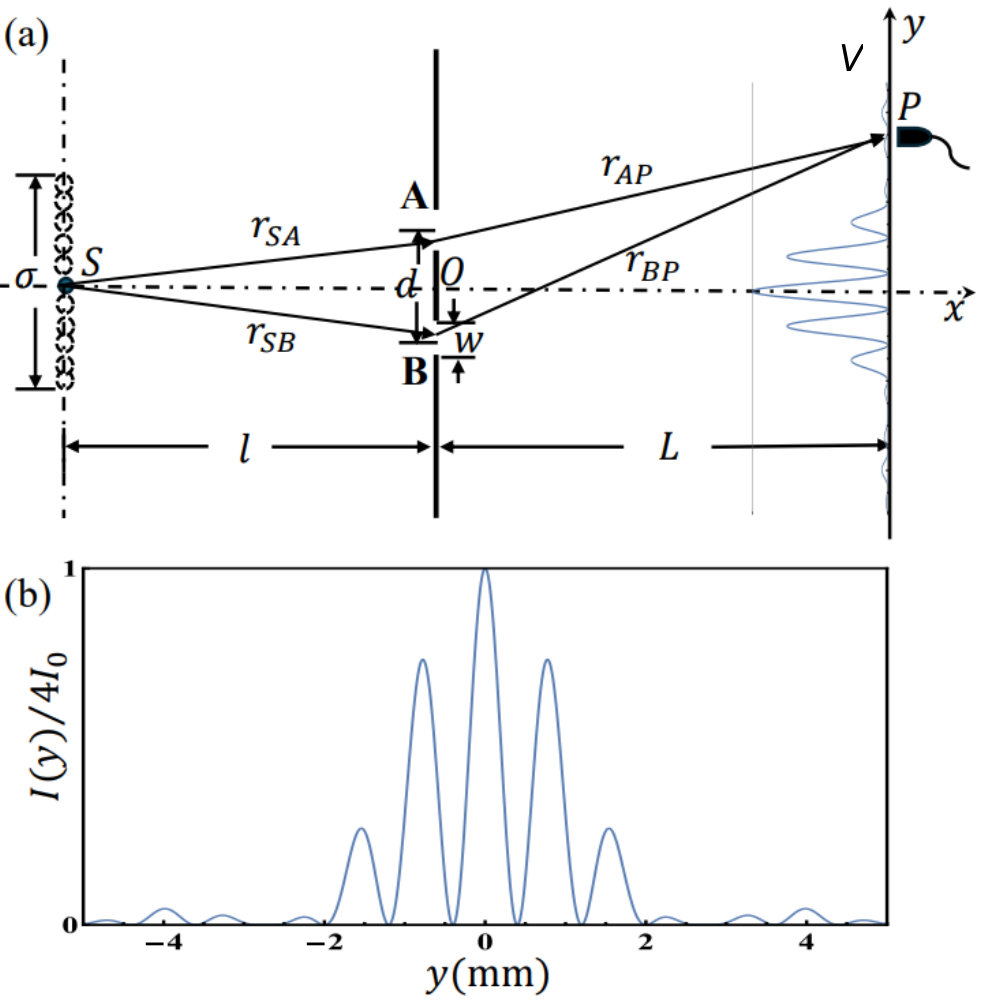}
\caption{(a) Schematic of the standard Young's experiment, showing that the diffraction-interference fringes and the source plane must be located on opposite sides of the double-slit plane. (b) Illustration of a typical first-order diffraction-interference pattern non-deterministically formed on the detection $y$-plane.}
\label{fig:standard}
\end{figure}

To facilitate our discussion, we begin by standardizing our notation with a brief overview of Young's experiment. In its classic setup, as depicted in Fig.~\ref{fig:standard}(a), two narrow slits, A and B, each with a width $w$ and separated by a distance $d$, are illuminated by a point monochromatic light source $S$ (solid circle). This point source emanates an optical field $E$ at wavelength $\lambda$ and wavenumber $k=2\pi/\lambda$ and is positioned at a distance $l$ along the optical $x$-axis from the origin $O$. The irradiance at point $P(L,y)$ on the observation screen $V$, located at a distance $L$ from the plane containing A and B, is determined by the superposition of the overall fields after two slits,
\begin{subequations}
\begin{align}
E_P=&E_A+E_B,\label{eq:yE}\\
E_A=&\frac{Ee^{ikr_{SA}}}{r_{SA}}\int^{\frac{d+w}{2}}_{\frac{d-w}{2}}ds\frac{e^{ikr_{AP}}}{r_{AP}},\label{eq:YEA}\\
E_B=&\frac{Ee^{ikr_{SB}}}{r_{SB}}\int^{-\frac{d-w}{2}}_{-\frac{d+w}{2}}ds\frac{e^{ikr_{BP}}}{r_{BP}}.\label{eq:YEB}
\end{align}
\end{subequations}
To simplify the discussion, hereafter we will concentrate on the case of the paraxial approximation with $r_{SA}=r_{SB}\simeq l$ and $r_{AP}=r_{BP}\simeq L+d^2/8L-ys/L$. Note that the optical path differentiation $ys/L$ is much smaller than $L$, so it can be disregarded in the amplitude factors in Eqs.~(\ref{eq:YEA}) and (\ref{eq:YEB}) to the lowest order. However, this path difference cannot be neglected in the phase factors. Consequently, the irradiance at $P(L,y)$ is computed by
\begin{equation}
I(y)=\frac{\epsilon_0c}{2}|E_P|^2=4I_0\mathrm{sinc}^2\left(\frac{\pi w}{\lambda L}y\right)\cos^2\left(\frac{\pi d}{\lambda L}y\right), \label{eq:Iy}
\end{equation}
where $I_0=\epsilon_0cw^2|E|^2/2l^2L^2$ with $c$ being the speed of light in vacuum and $\epsilon_0$ the permittivity of vacuum. This characteristic shape of Young's diffraction-interference intensity profile is illustrated in Fig.~\ref{fig:standard}(b), using a 500-nm cw laser to illuminate a double-slit with $w=0.15$ mm, $d=0.5$ mm, and $L=0.8$ m as an example. However, when an extended light source (dashed hollow circles) with a lateral dimension of $\sigma$ is introduced, the distinct interference fringes tend to blur and becomes diffused if $d\sigma/l>\lambda/2$, thereby establishing the spatial coherence criteria for an extended light source \cite{25}. 

Several critical conclusions can be readily drawn from Eq.~(\ref{eq:Iy}): (i) The resulting interference-diffraction structure, $I(y)$, is distributed `locally' on the observation screen $V$ and depends on the geometrical parameters $(L,y)$ extending from the double-slit plane to $V$, while remaining unaffected by the geometrical parameter $l$ from the source $S$ to the origin $O$. (ii) The extent of observable interference fringes is controlled by the diffraction effect originating from the slit aperture. (iii) The source $S$ and the observation screen $V$ are situated on opposite sides of the double-slit plane. (iv) Each point on the interference pattern emerges probabilistically or statistically from slit diffraction and cannot be chosen at will, making it unrealistic to have a one-to-one correspondence between the source and the observation. (v) The fundamental physics behind this phenomenon can be attributed to the behavior of individual particles (i.e., the single-particle effect), as verified by feeble light illumination \cite{2}. 

\section{Time-Reversed Young's Experiment}
\begin{figure}[htbp]
\centering\includegraphics[width=0.9\linewidth]{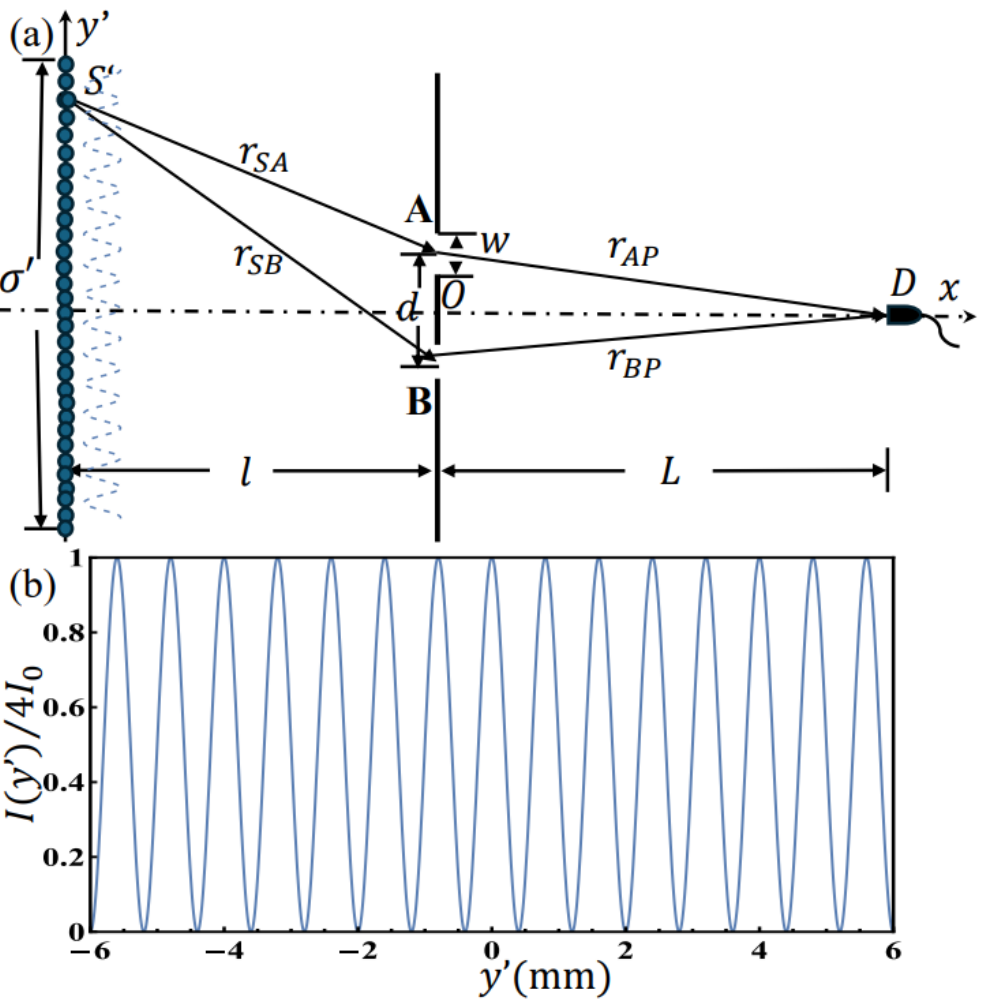}
\caption{(a) Schematic of the time-reversed Young's experiment, demonstrating that the nonlocal, nondiffractive interference fringes (dashed blue) and the source plane overlap and are located on the same side of the double-slit plane. (b) Illustration of a second-order diffractionless interference pattern (dashed blue line in (a)) deterministically formed by organizing the recorded data from a fixed-position detector $D$ based on the measured position coordinates of each individual active point light emitter on the $y'$-plane.}
\label{fig:reversed}
\end{figure}

However, as we delve into the analysis below, we find that each of the aforementioned conclusions can be systematically challenged through a time-reversal configuration of the typical Young's experiment, since this time-reversed operation leads to totally nonreciprocal physics. Also in sharp contrast, in the time-reversed scheme, there is no spatial coherence requirement on the lateral scale of the source whatsoever. Specifically, we are intrigued by a new rendition of Young's double-slit experiment, wherein we replace the original point source $S$ with a position-fixed point or bucket detector $D$, while concurrently substituting the observation screen $V$ with a laterally extended point light source $S'$, as schematic in Fig.~\ref{fig:reversed}(a). Through our forthcoming demonstration, we aim to highlight that despite both setups (Fig.~\ref{fig:standard}(a) versus Fig.~\ref{fig:reversed}(a)) adhering to time-symmetry operations, they inherently diverge and engender asymmetric, nonreciprocal physical effects. Unlike the first-order `local' diffraction-interference pattern on $V$ (Fig.~\ref{fig:standard}(b)), this time-reversed system invariably produces a `second-order nonlocal ghost' diffraction-free, clean interference fringes (Fig.~\ref{fig:reversed}(b)) determined by the information of both source and detection, devoid of the standard first-order interference effect. Here, we emphasize that the term `second-order interference' is not used in its conventional sense in quantum optics; its physical meaning and differentiation will become clear shortly.  

To clarify this concept, let us examine the new system schematic in Fig.~\ref{fig:reversed}(a) more carefully. Imagine a scenario with only one photo-detector, denoted as $D$, positioned at a fixed location $(L,0)$. In this scheme, $D$ only registers fluctuating light intensities or powers over time but lack the capability to discern the light's origin. Even when the recorded irradiance originates from two distinguishable paths, $D$ cannot differentiate between them. As a result, generating any meaningful pattern solely by the fixed-position $D$, including interference fringes, becomes unfeasible regardless of the spatial coherence of the light source. This observation highlights the necessity of extracting positional information from the light source to derive meaningful patterns. One potential solution is to ensure that only a single point emitter within the light source emits light at any moment, with its emission position being uniquely identifiable. How then, could we practically achieve such precise specifications?

Several strategies offer promise in this regard, providing avenues for experimenting the proposed concept: 

\underline{Solution I} (Sol-I): One possible way involves leveraging an ensemble of identical point light emitters, each capable of two-photon fluorescence. By ensuring that only one emitter fluoresces at any given moment, a photon from the emitted light can be directed to illuminate the double-slit aperture, while the other photon is utilized to map the emitter's position. This positional mapping could be achieved through a Gaussian thin lens imaging process, for example. Subsequently, by analyzing the photon trigger events detected by $D$, a nontrivial event distribution map can be post-generated, correlating with the positions of the individual point emitters. Note that the use of an optical lens will inevitably render the interference fringes---formed by the intensity pattern recorded by another detector corresponding to the position indicated by $D$---diffraction-limited in resolution.

\underline{Solution II} (Sol-II): Another more practical approach involves the development of a programmable light source. This entails the artificial construction of a uniform array of chromatic point light sources, where each emitter can be selectively activated to radiate light onto the double slits within a synchronized timeframe. Such on-demand activation may be realizable via different means. For instance, integrated electric fields can be utilized to sequentially excite each point source, with its spatial coordinate being simultaneously recorded. Alternatively, one can employ chemical markers or electro-optical effect to excite point emitters while simultaneously labeling their coordinate positions.

\underline{Solution III} (Sol-III): The third practical route is to fabricate a point source capable of precise spatial movement. To achieve this, for instance, one could affix a stable point light emitter, such as a quantum dot or an nitrogen-vacancy center in diamond, to the tip of a position-movable cantilever in such as atomic force microscopy. Then, the emitter can be laterally positioned with precision, maintaining a stationary position for an equal duration of emission at each point. As a consequence, the light intensities detected by the detector $D$ accurately mirror the spatial dynamics of the light source's movement.

It is evident that the essential objective across these three method categories is to establish a one-to-one correspondence between the captured light and the emitting source's precise spatial coordinates at that particular instance. This bit of spatial data information, crucial for the proposed time-reversed Young's experiment, enables the demonstration of diffractionless second-order interference by simply sorting the sequence of intensities or trigger events recorded on the position-fixed detector $D$. To facilitate this, we consider a point source $S'$ positioned at variable coordinates $(-l,y')$ along with different time moments in Fig.~\ref{fig:reversed}(a), emitting light onto the double slits. Subsequently, as the light traverses the double slits, it is intercepted by $D$ stationed at fixed coordinates $(L,0)$. The total electric field $E_D$ at $D$ now takes the form of
\begin{subequations}
\begin{align}
E_D=&E_A+E_B,\label{eq:ED}\\
E_A=&\frac{Ee^{ikr_{S'A}}}{r_{S'A}}\int^{\frac{d+w}{2}}_{\frac{d-w}{2}}ds\frac{e^{ikr_{AD}}}{r_{AD}},\label{eq:ETA}\\
E_B=&\frac{Ee^{ikr_{S'B}}}{r_{S'B}}\int^{-\frac{d-w}{2}}_{-\frac{d+w}{2}}ds\frac{e^{ikr_{BD}}}{r_{BD}}.\label{eq:ETB}
\end{align}
\end{subequations}
Again, under the paraxial approximation, we have $r_{S'A}\simeq l+d^2/8l-dy'/2l$, $r_{S'B}\simeq l+d^2/8l+dy'/2l$, and $r_{AD}=r_{BD}\simeq L$. Similarly, the optical path difference $dy'/l$ is negligible in the amplitude factors in Eqs.~(\ref{eq:ETA}) and (\ref{eq:ETB}) but cannot be neglected in the phase factors. The irradiance recorded at $D$ then assumes the following simpler result,
\begin{equation}
I(y')=\frac{\epsilon_0c}{2}|E_D|^2=4I_0\cos^2\left(\frac{\pi d}{\lambda l}y'\right),\label{eq:It}
\end{equation}
deterministic and ideal sinusoidal interference fringes that depend on the measured $y'$-position of the emitting light source plane and remain unaffected by diffraction at all! It is intriguing how Eq.~(\ref{eq:It}) appears impervious to the diffraction effect, even with the slit apertures in place. In addition, this intensity distribution is unequivocally dictated by the overall size $\sigma'$ of the (effective) light source. Thanks to the position-fixed detector $D$, the measurements remain effectively unaffected by slit diffraction--a challenge commonly encountered in previous studies, where the detector always captured the diffracted fields after passing through the aperture (Fig.~\ref{fig:standard}).  

In contrast to $I(y)$, $I(y')$ hinges on having instantaneous positional information of an active point light emitter, represented by a spatial correlation $\delta(y''-y')$ across the source plane. It is this spatial correlation in emission that defines the second-order correlation (or effective ``two-particle") effect, marking a fundamental departure from the single-particle perspective observed in standard Young's experiments using classical light, single photons, or single electrons. Alternatively, the $I(y')$ contour depends solely on the geometric properties from the source plane to the double-slit plane. Without the emitter's coordinate information, detector $D$ simply records a temporal sequence of light intensities or photon trigger events, lacking the first-order interference phenomenon. Although this differs from the standard definition of second-order correlation found in typical quantum optics textbooks---where it involves the product of intensities measured by two detectors---the `second-order' nature here arises because a position measurement is performed at the source rather than a regular intensity measurement. This position measurement is then \textit{correlated} with the intensity recorded by the positioned-fixed detector $D$. Another important distinction lies in the behavior of light in the classic Young's experiment: the light emerging from the slits is preserved (though diffracted), and is subsequently detected---randomly---at various positions on the observation plane $V$. In contrast, in the time-reversed Young's setup, the light collected by detector $D$ is not conserved in the same way, but instead clearly depends on the emission from the source plane $S'$. Furthermore, the interference pattern reconstructed by sorting intensities according to the measured emitter positions does not conform to the standard definition of first-order interference, which typically probes source spatial coherence within a well-defined coherent region. However, the time-reversed scenario surpasses this limitation and instead belongs to the domain of second-order correlations.

The construction of the interference fringes (\ref{eq:It}) is built upon organizing the recording data from $D$ according to the recorded positions of point emitters. As an example, consider an experiment in which a movable point source is turned on sequentially at positions $y'_1<y'_2<y'_3<y'_4<y'_5\cdots$, each for the same duration $\Delta t$. The corresponding powers recorded by a fixed detector $D$ are $P_1,P_2,P_3,P_4,P_5,\cdots$. The nonlocal, diffractionless interference pattern can then be obtained by plotting these power values against their respective $y'$. In fact, this is the method used to attain Fig.~\ref{fig:reversed}(b). This one-to-one correspondence enables a unique way of programming interference formations with the use of a programmable source like Sol-II and Sol-III as proposed above. Note that the ability to form such programmable interference fringes is exclusive to the time-reversed design is beyond the capabilities of traditional double-slit setups, regardless of whether classical or quantum (light) sources are employed. The emitter's spatial correlation within the source plane actually helps restore the time-reversal symmetry between detection and emission and hence erases their probabilistic relationship, which is however inherent in traditional arrangements due to the unavoidable diffraction effects behind the slits. From this perspective, the non-diffractive spatial interference fringes (\ref{eq:It}) can be seen as a \textit{spatial} analogue of a Mach-Zehnder interferometer \cite{26,27,28,29,30} with single photons. However, the underlying physics appears fundamentally different: the former arises from a second-order interference effect as explained above, whereas the latter is simply a first-order interference effect.


On the other hand, the proposed time-reversed version must display a certain degree of complementarity with the conventional setup. Indeed, this becomes evident when noting that both $I(y)$ and $I(y')$ share the same intensity normalization constant $I_0$. A comparison between Eqs.~(\ref{eq:YEA}) and (\ref{eq:YEB}) with Eqs.~(\ref{eq:ETA}) and (\ref{eq:ETB}) reveals that, although the spherical wavelets before the slits in Young's experiment do not directly contribute to interference pattern, they are nonetheless essential for its formation in the time-reversed case. In the traditional Young's experiment (Fig.~\ref{fig:standard}), the diffraction-interference effect originates precisely from the light beyond the slits, whereas in the time-reversed configuration, this post-slit propagation remains fixed. Moreover, if the fixed detector $D$ in the time-reversed setup were replaced by a movable observation screen $V$, one would recover a series of conventional, laterally shifted diffraction-interference patterns. These observations underscores the intrinsic complementarity between the two configurations. Such complementarity and distinction in a time-reversed framework may offer fresh insights into foundational aspects of quantum mechanics.

\section{Advanced-Wave Interpretation}
To have an alternative perspective on the defined second-order effect, here we present an advanced-wave pictorial description. In this conceptual depiction of time-reversal, a photon is imagined to originate from detector $D$ at the fixed position, propagate backward the source, and transfers its propagation details to a `second (virtual) photon.' This `second photon' is then considered to be `locally detected at the source position,' effectively retracing all possible trajectories of the originally transmitted photon. In this view, the source and detection planes are conceptually aligned, marking a significant departure from conventional interpretations. Importantly, spatial correlation at the source plays a critical role in suppressing diffraction effects during measurement. In our system, the resolution of the constructed pattern is not contingent upon light diffraction, but rather relies on the accurate measurement of the emitter's position. This key feature enables us to fundamentally surpass the Rayleigh diffraction limit, providing a powerful far-field strategy for super-resolution optical imaging and sensing---particularly relevant to biological and medical applications. It offers advantages over existing near-field techniques such as photon scanning tunneling microscopy \cite{31} and superlens \cite{32}, as well as far-field approaches like confocal microscopy \cite{33}, 4Pi microscope \cite{34}, structured-illumination microscopy \cite{35,36}, fluorescence microscopy \cite{37,38,39,40,41}, and methods rooted in quantum optics \cite{42,43,44}. 

One might wonder whether our time-reversed double-slit experiment bears resemblance to the famous \textit{ghost diffraction-interference experiments} \cite{45,46}, wherein one photon from a pair of entangled photons illuminates the double-slit while being detected by a position-fixed, pointlike photon counting detector. Simultaneously, the other photon freely propagates to a spatially scanning photon counting detector, leading to the creation of nonlocal two-photon diffraction-interference fringes via coincidence counts. Despite both experiments falling under second-order correlation and lacking first-order interference, they are completely nonequivalent. The latter relies on quantum entanglement, particularly momentum correlation, between paired photons, whereas the former depends on spatial correlation in photon emission. Furthermore, according to Klyshko's advanced-wave picture \cite{47}, the ghost interference experiment can be interpreted as one photon being generated at a detector, traveling backward to pass through the slits and the source, where it becomes the second photon, before moving forward in time to reach the second detector. This interpretation contrasts with our above explanation of the latter setup, where the `second (virtual) photon' does not necessitate propagation to generate the pattern. Moreover, the diffraction effect persists in the two-photon ghost interference experiment, but it is entirely absent in the time-reversed arrangement. All in all, the two types of experiments obey distinctive physical principles.

\section{Analysis on Phase Sensitivity and Spatial Resolution}
In addition to its fundamental implications, the time-reversed Young's experiment offers practical utility in precision metrology and superresolution imaging. A key question is: how small a displacement (transverse or longitudinal) of a point-like emitter can be reliably detected by analyzing the resulting interference pattern? To address this, we perform a detailed sensitivity and resolution analysis for both transverse and longitudinal perturbations.

We consider a point emitter located at a transverse position $y'$ and axial distance $l$ from a symmetric double-slit mask of slit separation $d$ (with $l \gg d$). The emitted light propagates through both slits and interferes at a fixed-position or a bucket detector, producing an intensity signal modulated by the phase difference accumulated due to geometric path imbalances. The detector is assumed to have unit quantum efficiency, and on average $\bar{N}$ photons are collected. We analyze the impact of small emitter displacements in the transverse ($\delta y'$) and longitudinal ($\delta l$) directions.

\subsection{Transverse Phase Sensitivity and Resolution}
We begin by examining the transverse phase sensitivity. When the emitter is displaced transversely by a small amount $\delta y'$ from a general location $y'$, the geometric path difference between the two slits changes, resulting in a phase shift
\begin{equation}
\Delta\phi_\perp(y')=\frac{2\pi d}{\lambda l}y', \;\text{so} \; 
\Delta\phi_\perp(y'+\delta y')=\Delta\phi_\perp(y')+\delta\phi_{\perp},
\end{equation}
where
\begin{equation}
\delta\phi_{\perp}= \frac{2\pi d}{\lambda l} \delta y'.
\end{equation}
The interference intensity at the detector becomes
\begin{align}
&I(y'+\delta y')=2I_0\left[1+\cos\left(\Delta\phi_\perp(y')+\delta\phi_{\perp}\right)\right]\nonumber \\
&\approx 2I_0\left[1+\cos(\Delta\phi_\perp)-\sin(\Delta\phi_\perp)\delta\phi_\perp-\frac{1}{2}\cos(\Delta\phi_\perp)\delta\phi^2_\perp\right],
\end{align}
where we used a Taylor expansion for small $\delta\phi_\perp$. To leading order, the change in intensity is
\begin{equation}
\delta I\approx-2I_0\sin\left(\frac{2\pi d}{\lambda l}y'\right)\cdot\frac{2\pi d}{\lambda l}\delta y'.
\end{equation}
Hence, the sensitivity is position-dependent:
\begin{equation}
\frac{dI}{dy'} = -2I_0 \cdot \frac{2\pi d}{\lambda l} \cdot \sin\left( \frac{2\pi d}{\lambda l} y' \right).
\end{equation}

The maximum sensitivity occurs when the sine function is $\pm 1$, i.e., at the quadrature points:
\begin{equation}
y' = \left(2n + 1\right) \frac{\lambda l}{4d}, \quad n \in \mathbb{Z}.
\end{equation}
At these positions, the slope becomes the steepest:
\begin{equation}
\left| \frac{dI}{dy'} \right|_{\text{max}} = 2I_0 \cdot \frac{2\pi d}{\lambda l}.
\end{equation}

Assuming shot-noise-limited detection, the intensity fluctuation is $\delta I_{\text{noise}} = \sqrt{2I_0}$. Therefore, the minimum resolvable transverse displacement is
\begin{equation}
\delta y'_{\min}=\frac{\delta I_{\text{noise}}}{\left|\frac{dI}{dy'}\right|_{\text{max}}}=\frac{\lambda l}{2\pi d\sqrt{\bar{N}}}\;\left(\text{or }\frac{2y'}{(2n+1)\pi\sqrt{\bar{N}}}\right).\label{eq:deltaymin}
\end{equation}
This result shows that the transverse resolution improves with increased slit separation $d$, decreased wavelength $\lambda$, and higher photon number $\bar{N}$. The dependence on $l$ reflects the fact that finer angular variations at the slits translate into smaller spatial shifts at the source plane.

\subsection{Longitudinal Phase Sensitivity and Resolution}
Next, we consider axial displacements of the emitter. The optical path lengths from the emitter to the two slits are:
\begin{equation}
L_\pm = \sqrt{(l+\delta l)^2 + \left(y' \pm \frac{d}{2}\right)^2}.
\end{equation}
To second order in small parameters, the difference in optical path lengths is
\begin{align}
\Delta L &= L_+ - L_- \approx \frac{d y'}{l} \left(1 - \frac{2 \delta l}{l} \right),
\end{align}
so the accumulated phase shift due to $\delta l$ is
\begin{equation}
\Delta\phi_\parallel(y', \delta l) = \frac{2\pi}{\lambda} \Delta L 
\approx \frac{2\pi d y'}{\lambda l} \left(1 - \frac{2 \delta l}{l} \right).
\end{equation}
The dependence of the interference signal on axial displacement is then:
\begin{equation}
I(\delta l)\approx2I_0\cos\left[\frac{2\pi dy'}{\lambda l}\left(1-\frac{2\delta l}{l}\right)\right].
\end{equation}
Let the axial phase shift be defined as
\begin{equation}
\delta\phi_\parallel=-\frac{4\pi dy'}{\lambda l^2}\delta l,
\end{equation}
and then the intensity variation becomes
\begin{equation}
\delta I \approx 2I_0 \cdot \sin\left( \frac{2\pi d y'}{\lambda l} \right) \cdot \frac{4\pi d y'}{\lambda l^2} \delta l.
\end{equation}

This gives the sensitivity:
\begin{equation}
\frac{dI}{d(\delta l)} = 2I_0 \cdot \frac{4\pi d y'}{\lambda l^2} \cdot \sin\left( \frac{2\pi d y'}{\lambda l} \right),
\end{equation}
which attains its maximum when
\begin{equation}
y' = \left(2n + 1\right) \frac{\lambda l}{4d}, \quad n \in \mathbb{Z}.
\end{equation}
Thus the maximum slope is
\begin{equation}
\left|\frac{dI}{d(\delta l)}\right|_{\text{max}}=2I_0\cdot\frac{4\pi dy'}{\lambda l^2},\;\text{with }y'=\left(2n+1\right)\frac{\lambda l}{4d}.
\end{equation}
Substituting this back in, the minimum resolvable longitudinal displacement is
\begin{equation}
\delta l_{\min} = \frac{\delta I_{\text{noise}}}{\left| \frac{dI}{d(\delta l)} \right|_{\text{max}}}
= \frac{\sqrt{2I_0}}{2I_0 \cdot \frac{4\pi d y'}{\lambda l^2}} 
= \frac{\lambda l^2}{2\pi d y' \sqrt{\bar{N}}}.
\end{equation}
Evaluated at the optimal condition \( y' = \left(2n + 1\right)\frac{\lambda l}{4d} \), this simplifies better axial resolution to:
\begin{equation}
\delta l_{\min}=\frac{2l}{(2n+1)\pi\sqrt{\bar{N}}}.\label{eq:deltalmin}
\end{equation}
This result show that smaller $\delta l_{\min}$ is achievable by increasing photon number $\bar{N}$, reducing emitter-to-slit distance $l$, or increasing quadrature index $n$, since axial resolution improves as $(2n+1)^{-1}$. However, this comes with a trade-off: Larger $n$ corresponds to larger transverse displacement $|y'|$. This may move the emitter outside the practical field of view or affect the system's response envelope (depending on the aperture, angular acceptance, or other physical constraints).

\subsection{Discussion}
The above results \eqref{eq:deltaymin} and \eqref{eq:deltalmin} highlight that both transverse and longitudinal spatial resolutions are ultimately set by the geometric configuration (slit separation $d$, source-slit distance $l$), the wavelength $\lambda$, and the mean photon number $\bar{N}$. Notably:

(i) Transverse resolution improves with larger slit spacing $d$ and shorter wavelengths.

(ii) Longitudinal resolution improves at fringe positions farther from the center ($y'\neq0$), where interference is more sensitive to optical path imbalance.

(iii) Both resolutions scale inversely with $\sqrt{\bar{N}}$, as expected under shot-noise-limited detection.

(iv) The use of general positions $y'$ rather than assuming $y'=0$ provides a more complete picture of how spatial resolution varies across the fringe field.

\noindent These findings have direct implications for the design of high-resolution imaging systems and precise emitter localization protocols based on time-reversed interferometric schemes. 

Unlike the classical Young's double-slit configuration, which suffers from diminished spatial resolution due to the diffraction envelope imposed by finite slit width $a$, the time-reversed approach avoids such limitations. In the traditional setup, the interference fringes are modulated by an envelope that decays away from the optical axis, reducing fringe visibility and limiting the maximum slope of the interference signal—thereby constraining sensitivity to small spatial displacements. By contrast, the time-reversed Young’s configuration exhibits phase sensitivity governed purely by the geometric path difference between the two arms. Crucially, it lacks an analogous diffraction envelope, allowing the signal to maintain high visibility across the ``virtual" detection plane. As a result, the scheme achieves significantly enhanced transverse and longitudinal resolution. This performance advantage stems from the intrinsic geometric phase accumulation in the emission process, rather than relying on intensity modulation patterns observed at the detection stage.

\section{Summary and Outlook}
To summarize, this theoretical proposal introduces a time-reversed double-slit experiment that inherently involves a nontrivial second-order correlation effect---though it departs conceptually from the standard definition in quantum optics. As argued throughout the text, the emergence of this \textit{nonlocal second-order interference} stems from a one-to-one correspondence between two distinct but correlated, spatially separated measurements: one that localizes the emitter's position at the source plane, and another that records the power at a fixed-position point detector. This correlation imbues the process with time-reversal symmetry and determinism in the construction of the interference pattern. 

Moreover, the scheme enables the digitized generation of programmable \cite{48} and deterministic, nondiffracting interference fringes, with their spatial extent determined by the lateral dimensions of an array of point emitters. The conceptual alignment of the source and image planes requires that only a single point emitter be active at any given time. Without knowledge of the emitter's position, the detector would register seemingly random intensity fluctuations. 

The analysis presented here assumes ideal conditions. Relaxing these assumptions will inevitably alter the resulting interference fringes. The sensitivity and resolution of these changes hold potential for applications in sensing and imaging---topics that will be addressed in future work. 

While light has served as the illustrative medium in this study, the same principle could, in principle, be extended to other substances such as electrons, atoms, and molecules. Given the broad relevance of superresolution imaging and sensing in modern science, we anticipate that this diffraction-free, spatial Mach-Zehnder interferometric scheme---characterized by a time-reversal-enforced one-to-one mapping between source and detection---could inspire new technological advances across a range of disciplines. 

Finally, although the phenomenon can be described classically, developing a complete quantum mechanical model of the process seems to be a nontrivial task.

\section*{Acknowledgement}
We are grateful to Min Xiao, Fengnian Xia, Qing Gu, Jiazhen Li, Hebin Li, Shengwang Du, Yanhua Zhai, and Saeid Vashahri Ghamsari for helpful discussions. This work was partially supported by the NSF ExpandQISE-2329027 and the DoE DE-SC0022069.

\section*{AUTHOR DECLARATIONS}
\textbf{Conflict of Interest}\\
The authors have no conflicts to disclose.

\textbf{Data availability}\\
The data that support the findings of this study are available from the corresponding author upon reasonable request.

\end{document}